\documentclass[10pt, aps,prc,twocolumn,superscriptaddress,preprintnumbers,
amsmath, 
floatfix,
longbibliography,
nofootinbib
]{revtex4-1}
\usepackage[T1]{fontenc}
\usepackage[utf8x]{inputenc} 
\usepackage{adjustbox}          
\usepackage[caption=false]{subfig}
\usepackage{url}
\usepackage{color}
\usepackage{float}
\usepackage[pdftex,colorlinks=true, linkcolor = blue, citecolor=blue,urlcolor=blue, bookmarksnumbered=true, bookmarksopen=true]{hyperref}
\usepackage{longtable}
\usepackage{amsfonts}
\usepackage{wrapfig,bm}
\usepackage[normalem]{ulem}
\usepackage{MnSymbol}
\newcommand{\beq}{\begin{equation}}
\newcommand{\eeq}{\end{equation}}
\newcommand{\bea}{\begin{eqnarray}}
\newcommand{\eea}{\end{eqnarray}}

\begin{document}

\title{
Pure quantum extension of the semiclassical Boltzmann-Uehling-Uhlenbeck equation}
  
\author{Aurel Bulgac }%
\email{bulgac@uw.edu}%
\affiliation{Department of Physics,%
  University of Washington, Seattle, Washington 98195--1560, USA}
  
\date{\today}

\begin{abstract}

The Boltzmann equation is the traditional framework in  which one  extends the time-dependent mean field classical description 
of a many-body system to include the effect of particle-particle collisions in an approximate manner. A semiclassical extension 
of this approach to quantum many-body systems was suggested by Uehling  and Uhlenbeck in 1933 for both 
Fermi and Bose statistics, and many further developments of this approach are known as the 
BoltzmannUehling-Uhlenbeck (BUU) equations. Here I introduce a pure quantum version of the BUU type of equations,
which is mathematically equivalent to a generalized Time-Dependent Density Functional Theory extended to superfluid systems.
As expected, during non-equilibrium processes the quantum Boltzmann one-body entropy increases during evolution.

\end{abstract}

\preprint{NT@UW-21-19}

\maketitle

\vspace{0.3cm}

The dynamics of a classical $N$-particle system can be described fully using the Liouville equation for the 
time-dependent probability distribution function $f_N({\bm q}_1\ldots {\bm q}_N,{\bm p}_1\dots {\bm p}_N,t)$, where 
${\bm q}_k, {\bm p}_k$ are the canonical coordinates and momenta of the particles and $k=1\ldots N$. Integrating over 
$N-s$ coordinates and momenta one can introduce the $s$-particle time-dependent probability distributions
$f_s({\bm q}_1\ldots {\bm q}_s,{\bm p}_1\dots {\bm p}_s,t)$  and derive 
the Bogoliubov-Born-Green-Kirkood-Yvon (BBGKY) hierarchy of equations~\cite{Huang:1987}.  
The lowest order approximation to the exact BBGKY hierarchy is the Vlasov equation for the one-particle
time-dependent probability distribution function $f({\bm q},{\bm p},t)$
\begin{align}
\frac{\partial f}{\partial t} +\frac{\bm p}{m}\cdot\frac{\partial f}{\partial {\bm q}} +{\bm F}\cdot\frac{\partial f}{\partial {\bm p}} =0, \label{eq:V}
\end{align}
where $m$ is the particle mass (assuming that all particles have the same mass) and ${\bm F}$ is the average force experienced 
by a particle from all the other particles
\begin{align} 
{\bm F}({\bm q}_k) = -\sum_{l\neq k}^N \int d {\bm q}_l d{\bm p}_lf({\bm q}_l,{\bm p}_l,t) \frac{ V(|{\bm q}_k-{\bm q}_l |)}{\partial {\bm q}_k},   
\end{align}
assuming only two-particle interactions.  One can show that in the semiclassical approximation the time-dependent Hartree-Fock 
equations reduce to the Vlasov equation Eq.~\eqref{eq:V}.   
Boltzmann had the key insight to add an additional collision integral  to this equation, 
assuming ''molecular chaos'' prior to the two particle collision, and thus arriving at a kinetic equation. \textcite{Nordheim:1928} and 
\textcite{Uehling:1933} generalized the Boltzmann equation by modifying the collision integral to take into account the quantum 
statistics, known as the BoltzmannUehling-Uhlenbeck (BUU) equation, and see also \textcite{Bertsch:1988} for applications to nuclear physics,
\begin{align}
&\frac{\partial f}{\partial t} +\frac{\bm p}{m}\cdot\frac{\partial f}{\partial {\bm q}} +{\bm F}\cdot\frac{\partial f}{\partial {\bm p}} =I_\text{coll}({\bm p},t), \label{eq:B}\\
&\!\!\! \!\!I_\text{coll}({\bm r},{\bm p},t) = -\frac{1}{(2\pi\hbar)^3}
\int \!\!d\Omega \int \!\!d {\bm p}_2 \int \!\!d{\bm p}_4\, v\frac{ d\sigma (q,\Omega )}{d \Omega}\label{eq:collq}\\
&\!\!\!\times  \{  f({\bm r},{\bm p},t) f({\bm r},{\bm p}_2,t) [1+\theta  f({\bm r},{\bm p}_3,t)] [1+\theta f({\bm r},{\bm p}_4,t)]\nonumber \\
&\!\!\! -f({\bm r},{\bm p}_3,t)f({\bm r},{\bm p}_4,t) [1+\theta f({\bm r},{\bm p},t)][1+\theta  f({\bm r},{\bm p}_2,t)]  \}, \nonumber \\
&\times \delta({\bm p} +{\bm p}_2 -{\bm p}_3-{\bm p}_4),\nonumber \\
& mv= q=|{\bm p}-{\bm p}_2|.
\end{align} 
Here $\theta = \pm 1$ for bosons and fermions respectively, and $\theta \equiv 0$  in the original Boltzmann equation.  
$\tfrac{d\sigma(q,\Omega)}{d\Omega}$ is the differential cross section of 
particles with initial or final ${\bm p}, {\bm p}_{2}$ and final/initial momenta ${\bm p}_{3,4}$ 
into a solid angle $d\Omega$. The integrals of the first and the second terms  in the curly brackets 
in Eq.~\eqref{eq:collq} are often referred as the loss and gain terms in this kinetic equation. 

The numerical solution of the BUU equation is significantly simpler than 
the solution of the time-dependent Hartree-Fock (TDHF) equations. For example, for a nuclear system 
in a simulation box $L^3=50^3$ fm$^3$ and with a momentum cutoff of $p_\text{cut}=600$ MeV/c there are 
$4L^3(2p_\text{cut})^3/(2\pi\hbar)^3\approx 4.56\times 10^5$ quantum phase-space cells, 
while for a TDHF solution a system of $N=500$ nucleons in the same volume $50^3$ fm$^3$ and with a spatial lattice constant of $l=1$ fm, 
which corresponds to the same momentum cutoff $p_\text{cut} =\pi\hbar/l\approx 600$ MeV/c, 
has a total of $2NL^3(2p_\text{cut})^3/(2\pi\hbar)^3\approx1.1\times 10^8$ quantum
phase-space cells. However, since collisions are absent in TDHF framework, the role of equilibration processes are severely underestimated,
even though TDHF describes more accurately the single-particle quantum dynamics  and operates in a bigger space.

Similarly to the original Boltzmann equation, the BUU equation is valid only for a quantum dilute weakly interacting system 
in the semiclassical approximation. Therefore the particle-particle interaction  has to be weak and short-ranged, and the average 
interparticle separation should be smaller than the interaction range.   
However, most of the quantum many-body systems of interest are dense, as the interaction range is of the order of the 
average interparticle separation or even larger, and the interaction strength is typically strong and in such situations the
evaluation of the collision integral relies on various approximations and assumptions, and their accuracy and/or validity 
is almost impossible to evaluate. In a dense system the use of the free space cross section $\tfrac{d\sigma(q,\Omega)}{d\omega}$ 
is highly questionable, $n$-body collisions with $n>2$ should be taken into account, and the assumption that a collision occurs at 
a well defined point in space ${\bm r}$ and the absence of memory effects are inconsistent with the quantum uncertainty principle.     

There were many attempts over the years to develop time-dependent descriptions of many-nucleon systems 
beyond the mean field, in order to describe missing two-body correlations, and in particular to allow for the
equilibration of the single-particle degrees of freedom, while at the same time aiming towards a correct description of the quantum single-particle dynamics. 
The earliest attempts can be traced back to the generator coordinate method (GCM) and its time-dependent extension suggested by
Wheeler and collaborators~\cite{Hill:1953,Griffin:1957}, see a recent review~\cite{Verriere:2020}. One can try to introduce 
explicitly the two-body densities as well, see the recent review~\cite{Toyama:2020}. Other authors  have suggested adding stochastic 
terms to the TDHF equations and I refer the interested reader to Ref.~\cite{Bulgac:2019a}, where a number of such approaches are discussed. 
It suffice to say that these attempts have limited success in practice for many-fermion systems,
apart from applications to rather idealized and simple cases. 

I present arguments that a generalization of the extension of the Time-Dependent Density Functional Theory (TDDFT) to superfluid systems 
is a generalized mean field framework, which can accommodate two body collisions. I use the acronym gTDDFT for this further generalization, 
which will be still local, in the spirit of  the Kohn-Sham approach~\cite{Kohn:1965fk} to the Density Functional Theory (DFT), 
often referred in literature as the local density approximation (LDA) or its further extensions~\cite{Dreizler:1990lr}.
The  DFT is  in principle mathematically 
equivalent with the many-body Schr{\"o}dinger equation at the level of one-body 
density~\cite{Hohenberg:1964,Kohn:1965fk,Dreizler:1990lr,Gross:2006,Gross:2012}.
The difficulties with both these quantum many-body approaches are well known. The Schr{\"o}dinger equation requires the nucleon-nucleon 
interactions, which are not known exactly, and for systems of many nucleons the numerical solution of this equation is practically 
impossible, unless various approximations are introduced. Within DFT one needs to know the energy density functional (EDF), 
which cannot be independently measured, its relation with the nucleon-nucleon interaction cannot be accurately established, 
and for time-dependent phenomena memory effects maybe  important~\cite{Gross:2006,Gross:2012}. 
The current difficulties of {\it ab initio} calculations and 
their relation with DFT approaches were recently discussed by \textcite{Salvioni:2020}. \\

\begin{figure}[t]
\includegraphics[width=0.9\columnwidth]{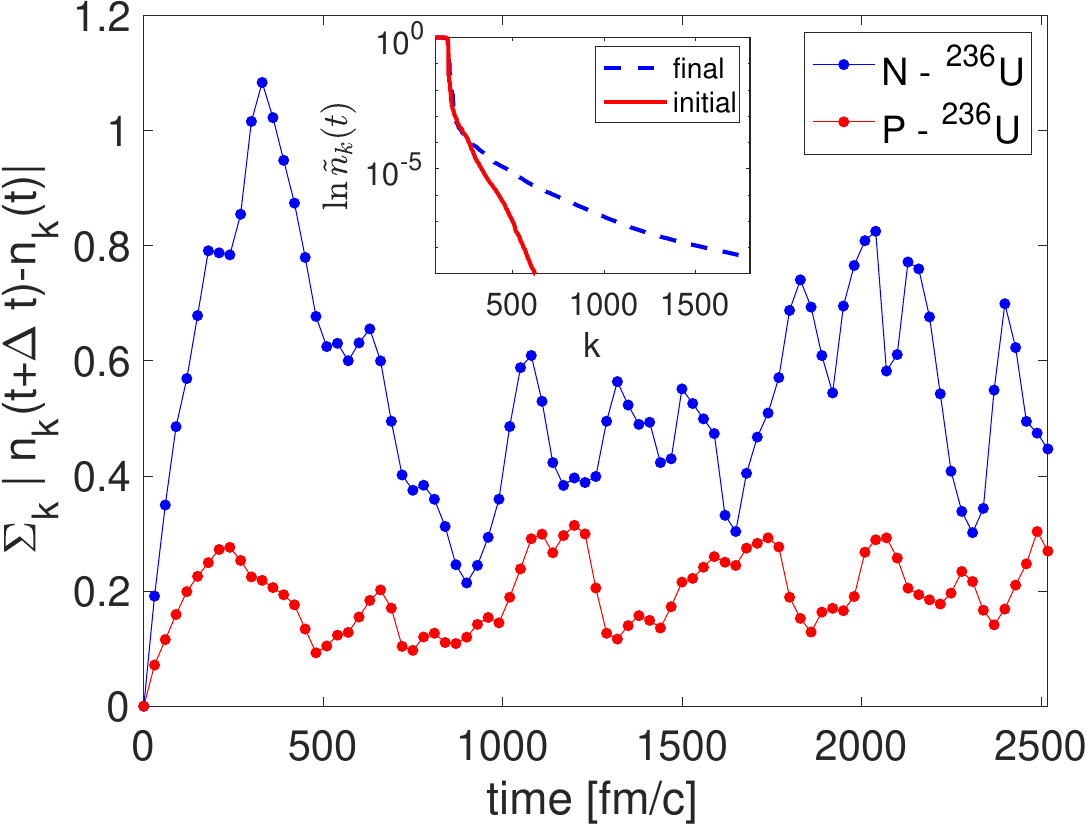}
\includegraphics[width=0.9\columnwidth]{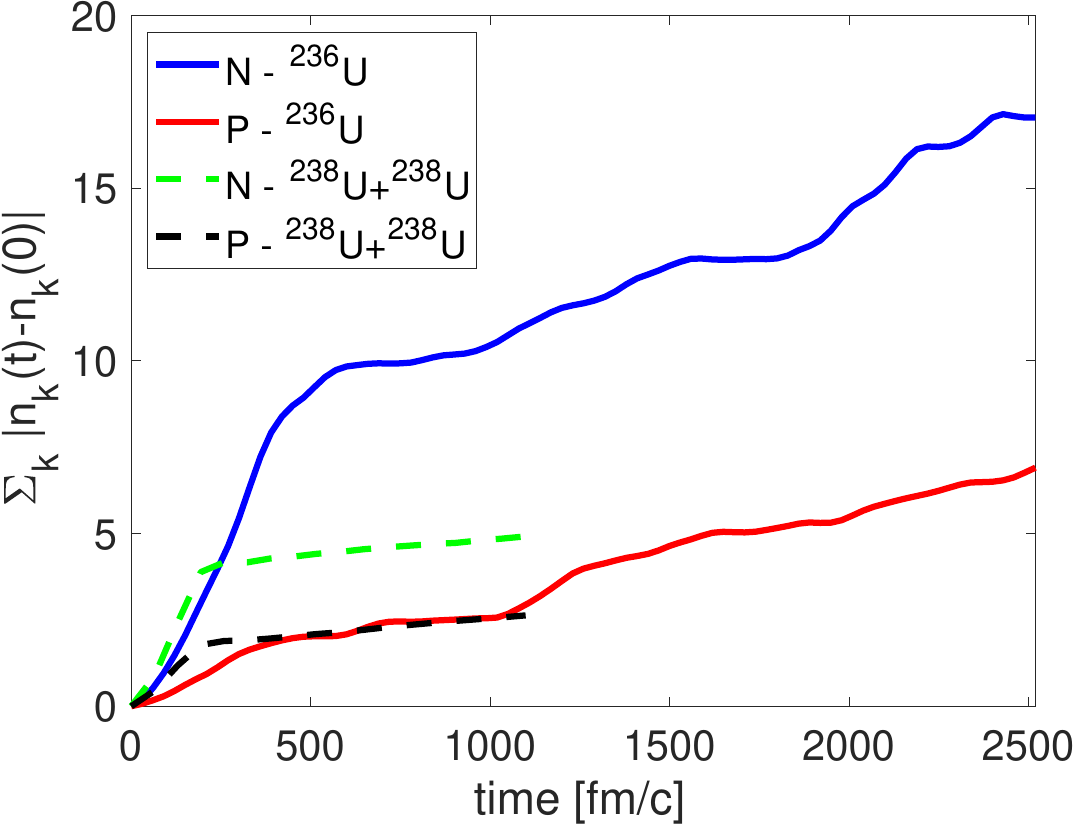}
\caption{ \label{fig:Fission} Typical time evolutions of the nucleon occupation probabilities
in a  TDSLDA simulation of induced fission of $^{236}$U started near the top of the outer fission 
barrier until complete fission fragment separation and for  XY-collision (see Ref.~\cite{Golabek:2009} for convention)  
at zero impact parameter
of $^{238}$U+$^{238}$U with 1500 MeV initial center of mass frame energy. 
The simulations were performed with  the nuclear EDF SeaLL1~\cite{Shi:2018} with the LISE code~\cite{Jin:2021}.
Scission occurs at $t\approx 2300$ fm/c. In collision the two final fragments are fully separated at $t>1000$ fm/c. 
In the upper panel displays the evolution of short-time evolution of the cumulative nucleon occupation probability 
$\sum_k|n_k(t+\Delta t)-n_k(t)|$ with $\Delta t\approx 30$ fm/c for fission and $\Delta t\approx 64$ fm/c for collisions. 
In the inset the canonical neutron occupation probabilities $\tilde{n}_k(t)$, ordered by size,  are displayed at the start and finish of the simulation
and one can clearly see the formation of the long momentum tails.   The total change 
in the nucleon occupation probability $\sum_k|n_k(t)-n_k(0)|$ as a function of time is shown in the lower panel. 
Note that for any $\Delta t$ in the absence of pairing $\sum_k|n_k(t+\Delta t)-n_k(t)|\equiv 0$. 
Here $n_{k}(t) = \sum_{\sigma=\uparrow,\downarrow}\int  d{\bm r} |v_k({\bm r},\tau,\sigma,t)|^2$, for either $\tau=n,p$, see Eq.~\eqref{eq:uv}.}
\end{figure}

The generalized TDDFT (gTDDFT), which is a further
extension of the TDDFT to superfluid systems~\cite{Bulgac:2007,Bulgac:2010,Bulgac:2011a,Bulgac:2013a,Bulgac:2019}, which apart from 
allowing to describe static and time-dependent superfluid systems, 
has the side-effect of describing a particular class of two-body collisions.
We often refer to the TDDFT extended to superfluid systems in the spirit of Kohn-Sham local density approximation 
DFT~\cite{Kohn:1965fk} as the time-dependent superfluid local density approximation (TDSLDA), which will become thus gTDSLDA accordingly. 
As Bertsch {\it et al.} initially suggested~\cite{Bertsch:1980,Barranco:1990,Bertsch:1997,Bertsch:2017},
while a nucleus adiabatically elongates during fission the single-particle energy levels display typically avoided crossings.  
The naive picture is that at such an avoided level crossing a Landau-Zener transition may occur. If a nucleon does not undergo
a transition it will stay on the up-sloping level and a vacancy below the highest occupied level 
(the Fermi level) will be created by the down-sloping level. That means 
the nucleus will acquire an intrinsic excitation energy with a volume character, since the local Fermi surface will cease to be spherically symmetric. 
The dynamics of nuclei at relatively low energies is that of an incompressible quantum fluid, and its evolution is 
dominated by the surface tension and the shape of the electric charge distribution mostly~\cite{Meitner:1939,Bohr:1939}, with significant 
corrections due to shell-effects~\cite{Strutinsky:1967,BRACK:1972}. After many such avoided level 
crossings the nucleus will acquire a volume excitation energy in case of Landau-Zener transitions, 
an evolution unexpected for an incompressible fluid. 
That is the main reason why within a TDHF description of fission nuclei fail to reach scission~\cite{Goddard:2015,Goddard:2016,Tanimura:2015,Scamps:2015} and 
the presence of the pairing correlations in TDSLDA proved to be the crucial lubricant~\cite{Bulgac:2016,Bulgac:2019b,Bulgac:2020},
as expected for a long-time~\cite{Bertsch:1980,Barranco:1990,Bertsch:1997,Bertsch:2017}.
Pairing correlations provide the mechanism for the nucleus to follow the dynamics of an incompressible fluid, 
where the volume energy component does not dramatically change. 
The single-particle levels are typically characterized by Kramers degeneracies and when a nucleus approaches a level crossing
two nucleons jump together as a ``Cooper pair'' and the nucleus remains ``cold.''  Such a transition is also Bose enhanced in the presence of a 
pairing condensate~\cite{Bulgac:2016,Bulgac:2019b,Bulgac:2020}. 
Because of the presence of pairing correlations in both neutron and proton systems within TDSLDA 
nuclei can easily  undergo fission, unlike in a TDHF framework, when the initial configuration 
is close to the outer fission barrier.
The evolution mechanism championed by \textcite{Bertsch:1980} however implies the 
presence of neutron and proton pairing condensates. On the other hand, the overwhelming experimental evidence is that 
the fission dynamics is not an adiabatic process, which is at odds with the prevailing microscopic approaches, based on the 
assumption of adiabaticity of the large amplitude collective motion~\cite{Ring:2004,Schunck:2016,Pomorski:2012,Verriere:2020}.
The fission fragments emerge with a significant total excitation energy, which is up to 20\% 
of the total $Q =M_\text{ini}c^2 -M^\text{H}c^2-M^\text{L}c^2$ of the reaction, where $M_\text{ini}, M^\text{H}$ and $M^\text{L}$ are the 
masses of the initial fissioning nucleus in case of spontaneous fission and of the ground states of the prompt fission fragments, 
and $c$ is the speed of light.  If on the way from saddle-to-scission the emerging fission fragments become hot the presence 
of neutron and/or proton pairing condensates becomes highly questionable along with the mechanism suggested in Ref.~\cite{Bertsch:1980}.

The TDSLDA is formulated in terms of Bogoliubov quasiparticle wave functions (QPWFS). 
The evolution of nucleon QPWFS is governed by the equations:
\begin{align}\label{eq:TDSLDA}
i\hbar \frac{\partial}{\partial t}
\begin{pmatrix}
u_{k\uparrow}  \\
u_{k\downarrow} \\
v_{k\uparrow} \\
v_{k\downarrow}
\end{pmatrix}
=
\begin{pmatrix}
h_{\uparrow \uparrow}  & h_{\uparrow \downarrow} & 0 & \Delta \\
h_{\downarrow \uparrow} & h_{\downarrow \downarrow} & -\Delta & 0 \\
0 & -\Delta^* &  -h^*_{\uparrow \uparrow}  & -h^*_{\uparrow \downarrow} \\
\Delta^* & 0 & -h^*_{\downarrow \uparrow} & -h^*_{\downarrow \downarrow} 
\end{pmatrix}
\begin{pmatrix}
u_{k\uparrow} \\
u_{k\downarrow} \\
v_{k\uparrow} \\
v_{k\downarrow}
\end{pmatrix},
\end{align}
where I have suppressed the spatial $\bm{r}$ and time coordinate $t$, and $k$  
labels the QPWFS (including the isospin) $[u_{k\sigma}(\bm{r},t), v_{k\sigma}(\bm{r},t)]$, 
with $\sigma = \; \uparrow, \downarrow$ the z-projection of the nucleon spin. 
The single-particle (sp) Hamiltonian $h_{\sigma\sigma'}(\bm{r}, t)$, and the 
pairing field $\Delta(\bm{r}, t)$ are functionals of various neutron and proton 
densities, which are computed from the QPWFS~\cite{Jin:2017,Jin:2021}.

Typical evolution of the nucleon occupation
probabilities in a TDSLDA are shown in Fig.~\ref{fig:Fission}, which is absent in any TDHF, where $\dot{n}_k(t)\equiv 0$. 
In case of fission the emerging fission fragments have an excitation energy of $\approx 20$ MeV each.
In the case of collision $^{238}$U+$^{238}$U the final fragments have excitation energies of about 400 and 600 MeV respectively and the 
distance of closest approach is reached at $\approx 250 $ fm/c, leading to a heavy fragment 
with $Z\approx123$ and $N \approx198$.  At these excitations energies the neutron 
and proton pairing ``gaps'' have also significant spatial variations, the long-range order~\cite{Yang:1962} is absent, 
and the pairing ``gaps'' have also significantly decreased in magnitude and 
the ``true'' pairing condensates are therefore absent. However, the effect of these 
pairing ``gaps'' on the nucleon wave functions $v_{k\uparrow}v_{l\downarrow} \leftrightarrow  u_{m\uparrow} u_{n\downarrow} $ is 
basically the quantum equivalent of the action of the collision term in Eq.~\eqref{eq:collq} $f_1f_2\leftrightarrow  (1-f_3)(1-f_4)$.
It is notable that the rate of the single-particle occupation probability redistribution shown in Fig.~\ref{fig:Fission} 
\begin{align}
\sum_k|\dot{n}_k(t)|\approx\text{const.} \quad \text{for} \quad t> t_0,
\end{align}
is fairly constant after some initial time, $t_0\approx 350$ fm/c in the case of fission and $t_0\approx 200$ fm/c in the case of heavy-ion collisions, 
even after the reaction fragments are spatially separated. This is expected, as the thermal equilibration is a slower process.
While $\sum_k\dot{n}_k(t)\equiv 0$ is always satisfied,
in the absence of pairing correlations an even stronger constraint is in effect, $\dot{n}_k(t)\equiv 0$ for all $k$'s. 
During these initial transitory times  $t<t_0$, nuclei start with well-defined $nn$ and $pp$ pairing condensates, when the rates of pair transitions 
are higher due to the Bose enhancement mechanism. Since in the case of heavy-ion collisions the excitation energies are higher, 
the magnitudes of the remnant pairing fields are smaller than in the case of fission. 
In the case of $^{236}$U fission  one can demonstrate that the quantum Boltzmann one-body entropy,
\begin{align}
\!\!\!\!\! S(t)=-\sum_k [ \tilde{n}_k(t)\ln \tilde{n}_k(t) + (1\!-\!\tilde{n}_k(t) )\ln (1\!-\!\tilde{n}_k(t) ) ],\label{eq:S}
\end{align}
changes from $S(t_\text{ini})=12.4$ to $S(t_\text{fin})=23.0$, and thus entropy increases as expected 
in a non-equilibrium evolution, where $\tilde{n}_k(t)$  are canonical occupation probabilities. Performing neutron and proton particle projection
of the fissioning nucleus at the initial and final times, as described in Ref.~\cite{Bulgac:2021a},
leads to $S(t_\text{ini})=10.1$ to $S(t_\text{fin})=18.8$ for $^{236}$U and $S(t_\text{ini})=10.2$ to $S(t_\text{fin})=19.9$ for $^{238}$Pu.
 
A simple qualitative argument, assuming that pairing condensates are present,  
was presented in Refs.~\cite{Bertsch:1980,Barranco:1990,Bertsch:1997,Bertsch:2017}. 
During the fissioning of an axially symmetric fissioning nucleus
in a TDHF framework the projection of the single-particle angular momentum is conserved. In the initial nucleus the maximum nucleon 
orbital angular momentum is $l_z\approx k_Fr_0A^{1/3}$, which is noticeably larger than the maximum orbital angular momentum in a fission 
fragment $l_z\approx k_Fr_0(A/2)^{1/3}$. Here $k_F\approx 1.35$ fm$^{-1}$ is the Fermi wave vector and $r_0=1.2$ fm. 
Within TDHF  the single-particle occupation probabilities are conserved and in the absence of an effective 
mechanism for redistribution of the single-particle occupation probabilities the waist of the fission fragments 
are artificially kept large as in the initial nucleus, instead of shrinking by $\approx 2^{-1/3}\approx 0.79$.
In an axially symmetric nucleus two nucleons with conjugate momenta can easily jump simultaneously if
a transition $(m,-m)\rightarrow (m',-m')$ is allowed. Such a transition is controlled by a two-body matrix element
$\langle m,-m|V|m',-m'\rangle$, which describes a $nn$ or $pp$ collision with the pair quantum numbers $L=S=0, T_z=\pm 1$.   
Therefore, as in the case
of Boltzmann equation, the pairing correlations  allow for $nn$  and $pp$ collisions, but only with $L=S=0, T_z=\pm 1$. 
However, unlike the Boltzmann 
equation, the TDSLDA also allows for the Bose enhancement of such transitions. 

The absence of $np$ pair jumps is a major difference with the role played by the collision integral in the BUU equation. In heavy nuclei 
the number of $np$ pairs is larger than the sum of the numbers of $nn$ and $pp$ pairs and it is hard to accept that their role 
could be neglected in fission for example, particularly in the absence of genuine $nn$ and $pp$ pairing condensates.
I will show here how one can generalize the TDSLDA 
to include $np$-collisions with pair quantum numbers $L=0, S= 0,1$.  
It is important to appreciate the fact that even if the long-range order of the pairing field/condensate
is lost,  these two-nucleon transitions survive 
at large excitation energies of the fissioning nucleus and in the fission fragments, 
which  emerge with an excitation energy $\approx 20$ MeV, corresponding to 
intrinsic temperatures $\approx 1$ MeV or higher, as illustrated in Fig.~\ref{fig:Fission}. At these 
excitation energies both neutron and proton ``pairing'' fields have no phase coherence anymore, which means that 
the nucleons in the ``Cooper pairs'' have finite center-of-mass momenta, which vary from point-to-point inside the nucleus,
and the pairing fields have large spatial 
variation of their magnitudes~\cite{Bulgac:2016,Bulgac:2019b,Bulgac:2020}. 
In  spite of that, the rate of the redistribution of the nucleon occupation probabilities does not diminish for $t>t_0$, 
see Fig.~\ref{fig:Fission}. The addition of 
$np$ pairing short-range correlations is going to play a significant role in definition of the mass and charges fission yields, 
similarly in heavy-ion collisions. 
In nuclear and cold atom physics, pairing is attributed to an attractive short-range interaction, which as a result leads to very long 
momentum tails of the nucleon occupation probabilities $n(k)\propto 1/k^4$, which at the same time are always present due to 
the presence of short-range correlations~\cite{Sartor:1980,Tan:2008} and have been recently unequivocally put in evidence 
in experiments~\cite{Hen:2014b}, particularly in the case $np$-pairs, which, as  I advocate here, are likely the most important ones in dynamics.  

I introduce generalized Bogoliubov quasiparticle $u$ and $v$ components and 
corresponding generalized fermionic quasiparticle creation and annihilation operators 
\begin{align}
&{u}_k(x) = u_k({\bm r},\tau,\sigma,t), \quad {v}_k(x) = v_k({\bm r},\tau,\sigma,t), \label{eq:uv} \\
&\alpha_k^\dagger = \sumint dx [ \text{u}_k(x)\psi^\dagger(x) + \text{v}_k(x)\psi(x)],\\
&\alpha_k = \sumint dx [ {v}_k^*(x)\psi^\dagger(x) + {u}_k^*(x)\psi(x)],\\
&\{ \alpha^\dagger_k,\alpha_l\} = \delta_{kl}, \quad \{\alpha_k,\alpha_l\} =0,\\
&\{\psi^\dagger(x),\psi(y)\} = \delta(x-y), \quad \{\psi(x),\psi(y)\} = 0,
\end{align}
where $\tau = n,p$ and $\sigma =\uparrow,\downarrow$ and $\sumint$ stands for integration of spatial coordinates 
and summation over spin and isospin degrees of freedom. These new quasiparticle operators do not necessarily have a well defined isospin 
quantum number, they mix the neutrons and protons in the same manner as the spin degrees of freedom were 
mixed in previous approaches. With these definitions of quasiparticle states and with the restriction that the relevant anomalous 
densities be local in space one has to introduce the following four different types in the case when only $L=0$ is allowed:
\begin{align}
&\kappa_\tau({\bm r}) = \sum_k v_k^*({\bm r},\tau,\downarrow)u_k({\bm r},\tau, \uparrow),\, \tau = n,p\\
&\kappa_0({\bm r}) = \sum_k v_k^*({\bm r},n,\downarrow)u_k({\bm r},p,\uparrow), \\
&\kappa_1({\bm r}) = \sum_k v_k^*({\bm r},n,\uparrow)u_k({\bm r},p,\uparrow),
\end{align}
 where $\alpha_k|\Phi\rangle =0$. Here $\kappa_{n,p}({\bm r})$ are the usual neutron and proton anomalous densities, while 
 $\kappa_0({\bm r})$ describes $pn$-pairs with $S_z=0$ and $\kappa_1({\bm r})$ describes $pn$-pairs with $S_z=\pm 1$.  
 $\kappa_0({\bm r})$ has exactly the same form as the anomalous density for the unitary Fermi gas, in which case $p$ 
 and $n$ would refer to atoms in different hyperfine states, which sometimes could be different atom species. In  $\kappa_1({\bm r})$ the roles 
 of spin and isospin are switched when compared with $\kappa_\tau({\bm r})$.
 The normal densities have a similar spin-isospin structure
 \begin{align}
&n_\tau({\bm r}) =   \sum_{k,\sigma} v_k^*({\bm r},\tau,\sigma)v_k({\bm r},\tau,\sigma),\\
&n_{np}({\bm r})   =   \sum_{k,\sigma} v_k^*({\bm r},n,\sigma)v_k({\bm r},p,\sigma),\\
&{\bm \sigma}_\tau({\bm r}) =   \sum_{k,\sigma,\sigma'} v_k^*({\bm r},\tau,\sigma){\bm \sigma}_{\sigma,\sigma'} v_k({\bm r},\tau,\sigma'),\\
&{\bm \sigma}_{np}({\bm r}) =   \sum_{k,\sigma,\sigma'} v_k^*({\bm r},n,\sigma){\bm \sigma}_{\sigma,\sigma'} v_k({\bm r},p,\sigma'),
\end{align} 
and where ${\bm \sigma}$ are Pauli matrices.
Other types of densities (density gradients, currents, etc.) are also needed~\cite{Bender:2003,Perlinska:2004}.
The gTDSLDA equations read in this case
\begin{align} \label{eq:tdslda}
i\hbar \frac{\partial}{\partial t}
\begin{pmatrix}
\text{u}_k(x,t)  \\
\text{v}_k(x,t) 
\end{pmatrix}
=
\begin{pmatrix}
H & \Delta  \\
\Delta^\dagger &  -H^*
\end{pmatrix}
\begin{pmatrix}
\text{u}_k(x,t) \\
\text{v}_k(x,t)
\end{pmatrix}, 
\end{align}
where $\text{u}_k(x,t)$ and $\text{v}_k(x,t)$ are four-column vectors~\eqref{eq:uv} and $H$ and $\Delta$ are 
$4\times4$ matrix operators with the structure
\begin{align}
H=
\begin{pmatrix}
h_{n\uparrow,n\uparrow}({\bm r})& h_{n\uparrow,n\downarrow}({\bm r})&h_{n\uparrow\,p\uparrow}({\bm r})& h_{n\uparrow,p\downarrow} ({\bm r})\\
h_{n\downarrow,n\uparrow}({\bm r})& h_{n\downarrow,n\downarrow}({\bm r})&h_{n\downarrow\,p\uparrow}({\bm r})& h_{n\downarrow,p\downarrow} ({\bm r})\\
h_{p\uparrow,n\uparrow}({\bm r})& h_{p\uparrow,n\downarrow}({\bm r})&h_{p\uparrow\,p\uparrow}({\bm r})& h_{p\uparrow,p\downarrow}({\bm r}) \\
h_{p\downarrow,n\uparrow}({\bm r})& h_{p\downarrow,n\downarrow}({\bm r})&h_{p\downarrow\,p\uparrow}({\bm r})& h_{p\downarrow,p\downarrow}({\bm r}) 
 \end{pmatrix} \label{eq:H}
 \end{align} 
and
\begin{align}
\Delta = 
\begin{pmatrix}
0& \Delta_n({\bm r})&  \Delta_1({\bm r})&\Delta_0({\bm r})\\
-\Delta_n({\bm r})& 0&-\Delta_0({\bm r})&\Delta_1({\bm r})\\
-\Delta_1({\bm r})& -\Delta_0({\bm r})& 0&\Delta_p({\bm r})\\
 \Delta_0({\bm r})& -\Delta_1({\bm r})&-\Delta_p({\bm r})&0
 \end{pmatrix}.\label{eq:D}
 \end{align} 
I did not include the chemical potentials in Eq.~\eqref{eq:tdslda}, as their presence is not necessary in the time-dependent formulation.
Equations~\eqref{eq:tdslda} are derived via an EDF, which should satisfy all the usual required symmetries. 
In particular  the number and anomalous  mixed neutron-proton densities can enter in such an 
EDF only as combinations $|\kappa_0({\bm r})|^2$,    
$|\kappa_1({\bm r})|^2$, $|n_{np}({\bm r})|^2$, and $|{\bm \sigma}_{np}({\bm r})|^2$, in order to satisfy isospin invariance.
One can then show that both average neutron and proton numbers are conserved separately.  
Moreover, the average number of neutrons and protons with either spin-up or spin-down is conserved as well, 
unless an external time-dependent time-odd one-body field is present.  
If one assumes isospin symmetry then the three anomalous densities $|\kappa_{n,p}({\bf r})|^2$ and $|\kappa_0({\bm r})|^2$ 
should appear in the EDF with the same coupling constant.  The absence of a dineutron bound state and existence of a deuteron 
suggests however that $np$-pairs with $S=1,T=0$ could be controlled by a stronger effective $s$-wave coupling constant than the pairing coupling 
constant for $S=0, T=1$ pairs~\cite{Bertsch:2010,Gezerlis:2011}. This $np$ interaction can be derived either by eliminating the tensor interaction using second order 
perturbation theory or an approach similar to in medium similarity renormalization group~\cite{Stroberg:2019}. A conclusive experimental evidence of the presence of a genuine 
neutron-proton pairing condensate in nuclear ground states is absent, 
with perhaps the exception of $N=Z$ nuclei and remains a matter of 
debate~\cite{Perlinska:2004,Bertsch:2010,Gezerlis:2011,Frauendorf:2014,Romero:2019,Cederwall:2020}.  
The extension of the present analysis to $L\ge 1$ pairs is straightforward, 
see Ref.~\cite{Perlinska:2004}. 

Fission or heavy-ion collisions 
of superfluid nuclei are typically started from states with vanishing mixed normal and anomalous 
densities, which will remain so during the entire time-dependent evolution in the absence of $np$ mixing. 
The neutron-proton pairing correlations can lead to a significant 
redistribution of single-particle occupation probabilities, similar to the role played by the collision 
integral in BUU simulations~\eqref{eq:collq}. 
As a simple example one can consider the case of a nucleus where both $nn$ and $pp$ pairing correlations are absent 
and include only $np$ pairing or short-range correlations or collisions using the magic nucleus $^{100}$Sn. In the TDHF+TDBCS 
approximation the time evolution equations have a canonical form by design~\cite{Scamps:2012},  
the occupation probabilities evolve according to
\begin{align}
i\hbar \frac{d n_k}{dt} = \Delta_k\kappa^*_k - \Delta^*_k\kappa_k, \quad i\hbar \frac{d \kappa _k}{dt}=\Delta_k(1-2n_k)
\end{align}
where now one couples a neutron state $k$ with spin-up with a proton state $k$ with spin-up in the case of $S=1$ for example,
thus inter-changing the roles of the spin and isospin. These equations have exactly the same structure 
as in the case of either $nn$- or $pp$-pairing correlations,  
but with a different content of the pairing field, which now will describe the jumps of $np$-pairs.
Similar to the  BUU equation, a condensate
is not needed to facilitate mass and charge transport. If the system is susceptible to develop wide
mass and charge distributions one can initially simply seed relatively small pairing fields 
$\Delta_{0,1}({\bm r})$ as in Ref.~\cite{Bulgac:2021c}, and with an initial excitation energy corresponding to a larger level density. 
The Boltzmann one-body entropy 
will grow with time from $S(t_\text{ini})=0$, driving the system towards the most probable outcomes, as expected in a nonequilibrium 
process. Another option is to treat the pairing fields as phenomenological 
inputs as in nuclear BUU simulations. Since the occupation redistribution mechanism described here is similar 
to that present in the BUU equation, there is likely no need to generate $np$ components of the mean field 
part of Eq.\eqref{eq:H}, which were never considered  in the BUU equation as far as I know. 
The "true" mean field $h_{np}$ components are never dominant
and since they will lead only to uncorrelated one-particle jumps their role is negligible. 

In conclusion, noticing that  the TDSLDA describes transitions of 
$nn$ and $pp$ pairs even in the absence of genuine pairing condensates 
I have presented an extension of the TDSLDA framework, here dubbed gTDSLDA to account for 
$nn$, $pp$, and $np$ collisions, in a manner similar to the semiclassical BUU equation. 
The collision integral in BUU equation accounts of the loss and gain processes 
\begin{align}
f(x_1)f(x_2)\leftrightarrow [1\!-\!f(x_3)][1\!-\!f(x_4)].
\end{align} 
Exactly the same  type of transitions are performed by the $nn$, $pp$, and $np $ pseudo-pairing fields where transitions 
of the type~\cite{Bertsch:1980,Barranco:1990,Bertsch:1997,Bertsch:2017}
\begin {align}
\!\!\!\!\!\! v_k({\bf r},\sigma_1,\tau_1)v_l({\bm r},\sigma_2,\tau_2) \leftrightarrow 
u_m({\bm r},\sigma_3,\tau_3)u_n({\bm r},\sigma_4,\tau_4)
\end{align}
are enabled. In BUU and gTDSLDA frameworks
transitions occur at the same position in space. The $nn$ and $pp$ pairs jumps have been shown to occur consistently in the past 
TDSLDA calculations~\cite{Stetcu:2011,Bulgac:2016,Bulgac:2019b,Bulgac:2020} and Fig.~\ref{fig:Fission}, both in the presence
of genuine pairing condensates as well in their absence and they lead to an increase of the Boltzmann one-body entropy [see Eq.~\eqref{eq:S}].
gTDDFT or gTDSLDA thus incorporates naturally both the long-range mean field 
effects and the short-range correlations between nucleons.

I thank I. Abdurrahman for generating the data for Fig.~\ref{fig:Fission}, G. F. Bertsch for a number of
comments, and the anonymous referee for a very good suggestion.
The funding from the US DOE, Office of Science, Grant No. DE-FG02-97ER41014 and also the support provided 
in part by NNSA cooperative agreement DE-NA0003841 is greatly appreciated. This research used resources of the Oak
Ridge Leadership Computing Facility, which is a U.S. DOE Office of
Science User Facility supported under Contract No. DE-AC05-00OR22725
and of the National Energy Research Scientific Computing Center, which
is supported by the Office of Science of the U.S. Department of Energy
under Contract No. DE-AC02-05CH11231.


\providecommand{\selectlanguage}[1]{}
\renewcommand{\selectlanguage}[1]{}

\bibliography{latest_fission}

\end{document}